\documentclass{scrartcl}

\RequirePackage{ifthen,calc}
\newsavebox{\ffbox}\newlength{\ffboxlen}
\newcommand{\todo}[1]{%
  {\sbox{\ffbox}{\textbf{TODO:}\ \textit{{#1}}\ \textbf{:ODOT}}
    \settowidth{\ffboxlen}{\usebox{\ffbox}}
		\addtolength{\ffboxlen}{-5mm}
    \ifthenelse{\ffboxlen>\linewidth}{%
      \noindent\marginpar{$>>>>$}\textbf{TODO:}\ \textit{{#1}}\ \textbf{:ODOT}\marginpar{$<<<<$}}{%
      \noindent\marginpar{$>><<$}\textbf{TODO:}\ \textit{{#1}}\ \textbf{:ODOT}}}}

\usepackage{array,xspace,multirow,hhline,tikz,tabularx,booktabs,fixltx2e}

\usepackage[round]{natbib}
\usepackage{amsmath,amssymb,amsfonts,amsthm}

\newtheorem{definition}{Definition}%
\newtheorem{theorem}{Theorem}%
\newtheorem{proposition}{Proposition}%
\newcommand{\thmref}[1]{Theorem~\ref{#1}}

\newcommand{\propref}[1]{Proposition~\ref{#1}}

\newcommand{\remref}[1]{Remark~\ref{#1}}

\newcommand{\ie}{i.e.,\xspace}
\newcommand{\eg}{e.g.,\xspace}

\usepackage{calrsfs}

\newcommand{\teq}[1][]{\ifthenelse{\equal{#1}{}}{\mathit{TEQ}}{\mathit{TEQ}(#1)}}
\newcommand{\me}[1][]{\ifthenelse{\equal{#1}{}}{\mathit{ME}}{\mathit{ME}(#1)}}
\newcommand{\mc}[1][]{\ifthenelse{\equal{#1}{}}{\mathit{MC}}{\mathit{MC}(#1)}}
\newcommand{\uc}[1][]{\ifthenelse{\equal{#1}{}}{\mathit{UC}}{\mathit{UC}(#1)}}
\newcommand{\ba}[1][]{\ifthenelse{\equal{#1}{}}{\mathit{BA}}{\mathit{BA}(#1)}}
\newcommand{\cnl}[1][]{\ifthenelse{\equal{#1}{}}{\mathit{CNL}}{\mathit{CNL}(#1)}}
\newcommand{\tc}[1][]{\ifthenelse{\equal{#1}{}}{\mathit{TC}}{\mathit{TC}(#1)}}
\newcommand{\co}[1][]{\ifthenelse{\equal{#1}{}}{\mathit{CO}}{\mathit{CO}(#1)}}
\newcommand{\bp}[1][]{\ifthenelse{\equal{#1}{}}{\mathit{BP}}{\mathit{BP}(#1)}}

\newcommand{\teqrel}[1][]{\ifthenelse{\equal{#1}{}}{\boldsymbol\rightarrow}{\teqrel_{#1}}}

\newcounter{remark}
\newenvironment{remark}[1][]{\refstepcounter{remark} \ifthenelse{\equal{#1}{}}{\noindent\textbf{Remark~\theremark. }}{\noindent \textbf{Remark~\theremark~(#1).}}}{\medskip}

\title{Set-Monotonicity Implies Kelly-Strategyproofness}

\author{Felix Brandt\\
Technische Universit\"at M\"unchen\\Munich, Germany\\
 \texttt{\small brandtf@in.tum.de}}

\date{}

\begin{document}

\maketitle

\begin{abstract}
This paper studies the strategic manipulation of set-valued social choice functions according to Kelly's preference extension, which prescribes that one set of alternatives is preferred to another if and only if all elements of the former are preferred to all elements of the latter. 
It is shown that \emph{set-monotonicity}---a new variant of Maskin-monotonicity---implies Kelly-strategyproofness in comprehensive subdomains of the linear domain. Interestingly, there are a handful of appealing Condorcet extensions---such as the \emph{top cycle}, the \emph{minimal covering set}, and the \emph{bipartisan set}---that satisfy set-monotonicity even in the unrestricted linear domain, thereby answering questions raised independently by \citet{Barb77a} and \citet{Kell77a}.
\end{abstract}

 \noindent\textbf{JEL Classifications Codes: }D71, C70

\section{Introduction}
\label{sec:intro}

One of the central results in microeconomic theory states that every non-trivial social choice function (SCF)---a function mapping individual preferences to a collective choice---is susceptible to strategic manipulation \citep{Gibb73a,Satt75a}.
However, the classic theorem by \citeauthor{Gibb73a} and \citeauthor{Satt75a} only applies to \emph{resolute}, \ie single-valued, SCFs. The notion of a resolute SCF is rather restricted and artificial.\footnote{For instance,
\citet{Gard76a} claims that ``[resoluteness] is a rather restrictive and unnatural assumption.'' In a similar vein,
\citet{Kell77a} writes that ``the Gibbard-Satterthwaite theorem [\dots]
uses an assumption of singlevaluedness which is unreasonable'' and \citet{Tayl05a} that ``If there is a weakness to the Gibbard-Satterthwaite theorem, it is the assumption that winners are unique.'' This sentiment is echoed by various other authors \citep[see, \eg][]{Barb77b,Feld79b,Band83a,Band83b,DuSc00a,Nehr00a,ChZh02a}.}
For example, consider a situation with two agents and two alternatives such that each agent prefers a different alternative. The problem is not that a resolute SCF has to pick a single alternative (which is a well-motivated practical requirement), but that it has to pick a single alternative \emph{based on the individual preferences alone} \citep[see also,][]{Kell77a}. As a consequence, resoluteness is at variance with elementary notions of fairness such as neutrality and anonymity. 

In order to remedy this shortcoming, \citet{Gibb77a} went on to characterize the class of strategyproof \emph{social decision schemes (SDSs)}, \ie aggregation functions that yield probability distributions over the set of alternatives rather than single alternatives \citep[see also][]{Gibb78a,Barb79a}. 
This class consists of rather degenerate SDSs and Gibbard's characterization is therefore often interpreted as another impossibility result.
However, Gibbard's theorem rests on unusually strong assumptions with respect to the agents' preferences. In contrast to the traditional setup in social choice theory, which typically only involves ordinal preferences, his result relies on the expected utility axioms of 
\citet{vNM47a}, and hence on the existence of linear utility functions, in order to compare lotteries over alternatives.

The gap between Gibbard and Satterthwaite's theorem for resolute SCFs and Gibbard's theorem for SDSs has been filled by a number of impossibility results with varying underlying notions of how to compare sets of alternatives with each other \citep[\eg][]{Gard76a,Barb77a,Barb77b,Kell77a,Feld79a,MaPa81a,Band82a,Band83b,DuSc00a,BDS01a,ChZh02a,Sato08a,Umez09a}, 
many of which are surveyed by \citet{Tayl05a} and \citet{Barb10a}. 
In this paper, we will be concerned with the one of the weakest (and therefore least controversial) preference extensions from alternatives to sets due to \citet{Kell77a}. According to this definition, a set of alternatives is weakly preferred to another set of alternatives if all elements of the former are weakly preferred to all elements of the latter. 
A nice aspect of Kelly's preference extension is that its underlying behavioral assumptions are quite minimalistic (which strengthens impossibility results), yet reasonable enough to motivate meaningful positive results. Kelly's extension models that the agents are complete unaware of the tie-breaking mechanism that is used to eventually pick a single alternative. The question pursued in this paper is whether this uncertainty can be exploited to achieve strategyproofness.

\citet{Barb77a} and \citet{Kell77a} have shown independently that all non-trivial SCFs that are rationalizable via a quasi-transitive relation are manipulable according to Kelly's extension.\footnote{\citet{Barb77a} actually uses an extension that is even weaker than that of \citet{Kell77a}.} However, it is it well-known that (quasi-transitive) rationalizability by itself is unduly restrictive \citep[see, \eg][]{MCSo72a}.
As a consequence, \citet{Kell77a} concludes his paper by contemplating that ``one plausible interpretation of such a theorem is that, rather than demonstrating the impossibility of reasonable strategy-proof social choice functions, it is part of a critique of the regularity [rationalizability] conditions'' and \citet{Barb77a} states that ``whether a nonrationalizable collective choice rule exists which is not manipulable and always leads to nonempty choices for nonempty finite issues is an open question.''
Also referring to nonrationalizable choice functions, \citet{Kell77a} writes: ``it is an open question how far nondictatorship can be strengthened in this sort of direction and still avoid impossibility results.''
The condition of rationalizability has been significantly weakened in subsequent impossibility results \citep{MaPa81a,Band82a,Band83b}. At the same time, it has been noted that more positive results can be obtained for antisymmetric (\ie linear) individual preferences. 
In particular, it was shown that the omninomination rule \citep{Gard76a}, the Pareto rule \citep{Feld79a,Feld79b}, the Condorcet rule \citep{Gard76a,Nehr00a}, and the top cycle \citep{MaPa81a,Band83a,SaZw10a} are strategyproof when preference are linear (see \remref{rem:smon} for more details about these SCFs). However, all these rules are very indecisive and the latter two may even return Pareto-dominated alternatives. 

In this paper, we propose a new variant of Maskin-monotonicity for set-valued SCFs called \emph{set-monotonicity} and show that all set-monotonic SCFs are strategyproof in sufficiently rich subdomains of the linear domain. This covers all of the positive results mentioned above and proves that some---much more discriminating---SCFs are strategyproof. 
 Set-monotonicity requires the invariance of choice sets under the weakening of unchosen alternatives and is satisfied by the omninomination rule, the Pareto rule, the Condorcet rule,  and a handful of appealing Condorcet extensions such as top cycle, the minimal covering set, and the bipartisan set. Since set-monotonicity coincides with Maskin-monotonicity in the context of resolute SCFs, this characterization can be seen as a set-valued generalization of the \citet{MuSa77a} theorem. 

 \citet{Nehr00a} has proved a similar extension of the Muller-Sattherthwaite theorem by showing that \emph{Maskin-monotonicity} implies strategyproofness of set-valued SCFs in a sense marginally weaker than that of Kelly (see also \remref{rem:weakprefs}).\footnote{According to \citeauthor{Nehr00a}'s definition, a manipulator is only better off if he \emph{strictly} prefers all alternatives in the new choice set to all alternatives in the original choice set. For linear preferences, the two definitions only differ in whether there can be a single alternative at the intersection of both choice sets or not.} 
However, while Maskin-monotonicity is prohibitive in the general domain, set-monotonicity is not. The conditions themselves are independent, but we show that Maskin-montonicity implies set-monotonicity when assuming independence of unchosen alternatives.

We conclude the paper with a number of remarks concerning group-strategyproofness, stronger preference extensions, weak preferences, weaker domain conditions, and strategic abstention.

\section{Preliminaries}


Let $N=\{1,\dots,n\}$ be a finite set of agents, $A$ a finite and nonempty set of alternatives, and $\mathcal{L}$ the set of all linear (\ie complete, transitive, and antisymmetric) preference relations over~$A$. For $R_i\in \mathcal{L}$, $x \mathrel{R_i} y$ denotes that agent~$i$ values alternative~$x$ at least as much as alternative~$y$. We write~$P_i$ for the strict part of~$R_i$, \ie~$x \mathrel{P_i} y$ if~$x \mathrel{R_i} y$ but not~$y \mathrel{R_i} x$. The (strict) lower contour set of alternative $x$ with respect to $P_i$ is denoted by
$L(x,R_i)=\{y\in A \colon x\mathrel{P_i} y\}$.
For convenience, we will represent preference relations as comma-separated lists. For example, $a \mathrel{P_i} b \mathrel{P_i} c$ will be written as
$R_i \colon a, b, c$.
Two distinct alternatives $x$ and $y$ are \emph{adjacent} in $R_i$ if there is no $z$ with $x\mathrel{P_i} z \mathrel{P_i} y$. 


A (Cartesian) domain of preference profiles $\mathcal{D}$ is defined as $\mathcal{D}=\prod_{i\in N} \mathcal{D}_i \subseteq \mathcal{L}^N$. The maximal domain $\mathcal{L}^N$ will be referred to as the \emph{general domain}. 
We say that $R''$ lies in the \emph{comprehensive closure} of $R$ and $R'$ if for all $i\in N$, $R_i\cap R'_i\subseteq R''_i$. A domain $\mathcal{D}$ is \emph{comprehensive} if for all $R,R'\in \mathcal{D}$ and $R''\in \mathcal{L}^N$ such that $R''$ lies in the comprehensive closure of $R$ and $R'$, $R''\in \mathcal{D}$ \citep{Nehr00a}. For example, the domain of all linear extensions of a fixed partial order is comprehensive.
For a given preference profile $R\in\mathcal{D}$, $R_{-i}=(R_1,\dots,R_{i-1},R_{i+1},\dots,R_n)$ denotes the vector of all preference relations except that of agent $i$. An alternative $x\in A$ is called a \emph{Condorcet winner} if $|\{i\in N\colon x \mathrel{P_i} y\}|> n/2$ for all $y\in A\setminus \{x\}$.

Our central object of study are social choice functions. A \emph{social choice function (SCF)} is a function $f$ that maps a preference profile $R\in \mathcal{D}$ to a nonempty subset of alternatives $f(R)$.~$f$ is \emph{resolute} if $|f(R)|=1$ for all $R\in \mathcal{D}$.
A \emph{Condorcet extension} is an SCF that uniquely selects a Condorcet winner whenever one exists.

\subsection{Monotonicity}

We will consider three variants of monotonicity: a weak standard notion and two strengthenings, one of which was proposed by \citet{Mask99a} and one of which is new to this paper.
For a given preference profile $R$, an agent $i$, and two adjacent alternatives $x,y$ such that $y\mathrel{P_i} x$, $R^{i:(x,y)}$
denotes the preference profile in which agent $i$ swapped alternatives $x$ and $y$ and that is otherwise identical to $R$.
\begin{definition}\label{def:mon}
Let $R,R'\in \mathcal{D}$, $i\in N$, and $x,y\in A$ such that $R'= R^{i:(x,y)}$. Then, SCF~$f$ satisfies monotonicity, Maskin-monotonicity, or set-monotonicity, if
\begin{align*}
x\in f(R) &\text{ implies } x\in f(R')\text{,} \tag{monotonicity}\\
z\in f(R) \text{ and } y\in A\setminus\{z\} &\text{ implies } z\in f(R')\text{, or} \tag{Maskin-monotonicity}\\
Z=f(R) \text{ and } y\in A\setminus Z &\text{ implies } Z=f(R')\text{, respectively.} \tag{set-monotonicity}
\end{align*}
\end{definition}

The intuitive meaning of these definitions is as follows.
An SCF satisfies monotonicity if a chosen alternative remains in the choice set when it is strengthened with respect to another alternative;
it satisfies Maskin-monotonicity if a chosen alternative remains in the choice set when weakening another alternative; 
and it satisfies set-monotonicity if the choice set is invariant under the weakening of unchosen alternatives.\footnote{\citet{SaZw10a} study monotonicity properties for set-valued SCFs in general. None of the properties they consider is equivalent to set-monotonicity.}
\footnote{Note that set-monotonicity is in conflict with decisiveness. For instance, non-trivial set-monotonic SCFs cannot satisfy the (rather strong) positive responsiveness condition introduced by \citet{Barb77b}.}

Clearly, Maskin-monotonicity implies monotonicity and, as will be shown in \propref{pro:setmoniua}, set-monotonicity also implies monotonicity.
Despite the similar appearance, set-monotonicity is logically independent of Maskin-monotonicity.
Set-monotonicity has a stronger antecedent and a stronger consequence.
For example, consider a comprehensive single-agent domain consisting of the relations $R_1$ and $R'_1$ given by
\[
\begin{array}{rl}
R_1 \colon & a, b, c \text{, } \quad \text{and}\\
 R'_1 \colon &  b, a, c\text{.}
\end{array}
\]
If we define SCF $f$ by letting $f((R_1))=\{c\}$ and $f((R'_1))=\{b,c\}$, $f$ satisfies Maskin-monotonicity, but violates set-monotonicity.
If, on the other hand, we define $f$ by letting $f((R_1))=\{a,b,c\}$ and $f((R'_1))=\{b,a\}$, $f$ satisfies set-monotonicity, but violates Maskin-monotonicity.

Set-monotonicity coincides with Maskin-monotonicity in the context of resolute SCFs.  
Under the condition of \emph{independence of unchosen alternatives} which is satisfied by various set-valued SCFs, set-monotonicity is weaker than Maskin-monotonicity. Moreover, set-monotonicity implies independence of unchosen alternatives.
Independence of unchosen alternatives was introduced by \citet{Lasl97a} in the context of tournament solutions (as ``independence of the losers'') and requires that the choice set is invariant under modifications of the preference profile with respect to unchosen alternatives.

\begin{definition}
An SCF~$f$ satisfies \emph{independence of unchosen alternatives} (IUA) if for all $R,R'\in\mathcal{D}$ such that $R_i|_{\{x,y\}}=R'_i|_{\{x,y\}}$ for all $x\in f(R)$, $y\in A$, and $i\in N$, $f(R)=f(R')$.
\end{definition}

\begin{proposition}
Maskin-monotonicity and IUA imply set-monotonicity.
\end{proposition}
\begin{proof}
Let~$f$ be an SCF, $R\in \mathcal{D}$, $i\in N$, $x\in A$, $y\in A\setminus f(R)$, and $R'=R^{i:(x,y)}$.
Maskin-monotonicity implies that $f(R)\subseteq f(R')$. Now, assume for contradiction that there is some $x'\in f(R')\setminus f(R)$. Since $x'\not\in f(R)$, it follows from Maskin-monotonicity that there is some  $y'\in A$ that is strengthened with respect to $x'$ when moving from $R'$ to $R$. Hence, $R'= R^{i:(x',y')}$, $x'=x$, and $y'=y$. Since $x=x'\not\in f(R)$ and $y\not\in f(R)$ by assumption, IUA implies that $f(R)=f(R')$, a contradiction.
\end{proof}

\begin{proposition}\label{pro:setmoniua}
Set-monotonicity implies monotonicity and IUA.
\end{proposition}
\begin{proof}
We first show that set-monotonicity implies monotonicity.
Let~$f$ be a set-monotonic SCF, $R\in\mathcal{D}$, $i\in N$, $x\in f(R)$, $y\in A$, and $R'\in R^{i:(x,y)}$. Clearly, in case $y\not\in f(R)$, set-monotonicity implies that $f(R')=f(R)$ and thus $x\in f(R')$. If, on the other hand, $y\in f(R)$, assume for contradiction that $x\not\in f(R')$. 
When moving from~$R'$ to~$R$, $y$ is strengthened with respect to outside alternative~$x$, and set-monotonicity again implies that $f(R)=f(R')$, a contradiction.
The fact that set-monotonicity implies IUA is straightforward from the definitions.
\end{proof}

\subsection{Strategyproofness}

An SCF is manipulable if an agent can misrepresent his preferences in order to obtain a more preferred outcome. Whether one choice set is preferred to another depends on how the preferences over individual alternatives are to be extended to sets of alternatives. In the absence of information about the tie-breaking mechanism that eventually picks a single alternative from any choice set, preferences over choice sets are obtained by the conservative extension~$\widehat R_i$ \citep{Kell77a}, where for any pair of nonempty sets $X,Y\subseteq A$ and preference relation $R_i$, 
\[X \mathrel{\widehat R_i} Y \text{ if and only if } x \mathrel{R_i} y\text{ for all }x\in X\text{ and }y\in Y\text{.}\]
Clearly, in all but the simplest cases, $\widehat R_i$ is incomplete, \ie many pairs of choice sets are incomparable. 
The strict part of $\widehat R_i$ is denoted by $\widehat P_i$, \ie $X \mathrel{\widehat P_i} Y$ if and only if $X \mathrel{\widehat R_i} Y$ and $x \mathrel{P_i} y$ for at least one pair of~$x\in X$ and~$y\in Y$. For linear preferences, $X \mathrel{\widehat P_i} Y$ if and only if $x \mathrel{P_i} y$ for all $x\in X$ and $y\in Y$ with $x\ne y$. Hence, $|X\cap Y|\le 1$.

\begin{definition}
An SCF is \emph{Kelly-strategyproof} if there exist no $R, R'\in\mathcal{D}$ and $i\in N$ with $R_{-i}=R'_{-i}$ such that $f(R') \mathrel{\widehat P_i} f(R)$.
\end{definition}

Kelly-strategyproofness is a very weak notion of strategyproofness. Nevertheless, most well-known SCFs such as \emph{plurality}, \emph{Borda's rule}, \emph{Copeland's rule}, \emph{Slater's rule}, or \emph{plurality with runoff}
fail to be Kelly-strategyproof in the general domain~\citep[see, \eg][Theorem 2.2.2]{Tayl05a}.

\citet{MuSa77a} have shown that, in the general domain, a resolute SCF is strategyproof if and only if it satisfies Maskin-monotonicity. Unfortunately, as famously shown by \citet{Gibb73a} and \citet{Satt75a}, only dictatorial or imposing resolute SCFs satisfy Maskin-monotonicity in the general domain. However, Maskin-monotonicity still implies strategyproofness of resolute SCFs in many restricted domains of interest \citep[see, \eg][]{KlBo13a}, including the class of comprehensive domains considered in this paper \citep{Nehr00a}.

\section{The Result}

\begin{theorem}\label{thm:main}
Every set-monotonic SCF on a comprehensive domain is Kelly-strategyproof.
\end{theorem}
\begin{proof}
Let $f$ be a set-monotonic SCF and $\mathcal{D}$ a comprehensive domain. 
We first show that set-monotonicity is equivalent to a version of set-monotonicity that is not restricted to pairwise swaps of adjacent alternatives. Rather, we require that choice sets may only change if an alternative is removed from a lower contour set of a chosen alternative.
 For $R,R'\in\mathcal{D}$ and $i\in N$ with $R_{-i}=R'_{-i}$, we say that $R'$ is an \emph{$f$-improvement over $R$} if for all $x\in f(R)$, $L(x,P_i)\subseteq L(x,P'_i)$. We claim that if $R'$ is an $f$-improvement over $R$, then $f(R')=f(R)$.
The statement can be shown by induction on $d(R,R')=|R_i\setminus R'_i|$. The induction basis is trivially satisfied because $R=R'$ if $d(R,R')=0$.
Assume the statement is true for all $R$ and $R'$ with $d(R,R')< k$ and consider $R$ and $R'$ such that $d(R,R')=k$. Since $R\ne R'$, there have to be two alternatives $x,y\in A$ such that $x\mathrel{P_i} y$ and $y\mathrel{P'_i} x$. Due to the transitivity of $R_i$ and $R'_i$, we may furthermore assume that $x$ and $y$ are adjacent in $R_i$. (However, $x$ and $y$ need not be adjacent in $R'_i$.)
$x\not\in f(R)$ because otherwise $R'$ is not an $f$-improvement over $R$. Set-monotonicity implies that $f(R^{i:(y,x)})=f(R)$ and comprehensiveness that $R^{i:(y,x)}\in \mathcal{D}$. It follows from $d(R^{i:(y,x)},R')<k$ and the induction hypothesis that $f(R^{i:(y,x)})=f(R')$. Hence, $f(R)=f(R')$,


Now, for the proof of the statement of the theorem, assume for contradiction that $f$ is not Kelly-strategyproof. Then, there have to be $R,R'\in\mathcal{D}$, and $i\in N$ with $R_{-i}=R'_{-i}$ such that $f(R')\mathrel{\widehat{P}_{i}} f(R)$. The latter obviously entails that $f(R)\ne f(R')$.
The proof idea is to find some $R^*$ in the comprehensive closure of $R$ and $R'$ (and hence in $\mathcal{D}$) such that $R^*$ is an $f$-improvement over both $R$ and $R'$, which implies that $f(R)=f(R')$, a contradiction.

Let $m\in f(R')$ such that $L(m,P_i)\cap f(R')=\emptyset$.
In other words, $m$ is the alternative in $f(R')$ that is ranked lowest in $R_i$. Note that, if $f(R)\cap f(R')\ne \emptyset$, then $f(R)\cap f(R') = \{m\}$. Next, we partition $A$ into the strict lower contour set and the upper contour set of $m$ with respect to $R_i$, \ie 
$L = L(m,P_i)$ and $U=A\setminus L$ and
define a new preference profile by letting
\[
R^*_i = R_i|_L \cup R'_i|_U \cup \{(x,y) \colon x\in U, y\in L\}\text{,}
\]
\ie the upper part of $R_i^*$ is ranked as in $R_i'$ and the lower part as in $R_i$, and $R^*=(R_{-i},R_i^*)=(R'_{-i},R_i^*)$.

$R^*$ lies in the comprehensive closure of $R$ and $R'$ (and hence in $\mathcal{D}$) because for all $x,y\in A$ with $x\mathrel{R_i^*} y$, we have $x\mathrel{R_i} y$ or $x\mathrel{R'_i} y$. For $x,y\in U$, this follows from $R^*_i|_U = R'_i|_U$; for $x,y\in L$, from $R^*_i|_L = R_i|_L$; and for $x\in U, y\in L$ from $(x,y)\in R_i$.

$R^*$ is an $f$-improvement over $R$ because $f(R)\subseteq L\cup \{m\}$ and $L(x,P^*_i)=L(x,P_i)$ for all $x\in L$ and $L(m,P^*_i)\supseteq L(m,P_i)=L$. Set-monotonicity then implies that $f(R^*)=f(R)$.
$R^*$ is an $f$-improvement over $R'$ because $f(R')\subseteq U$ and $L(x,P^*_i)=L(x,P'_i)\cup L$ for all $x\in U$. Hence, set-monotonicity implies that $f(R^*)=f(R')\ne f(R)$, a contradiction.
\end{proof}

We conclude the paper with nine remarks.
\medskip

\begin{remark}[Set-monotonic SCFs] \label{rem:smon}
There are a number of rather attractive SCFs that satisfy set-monotonicity in the general domain $\mathcal{L}^N$. In particular, every monotonic SCF that satisfies the \emph{strong superset property (SSP)} (\ie choice sets are invariant under the removal of unchosen alternatives) also satisfies set-monotonicity.\footnote{The strong superset property goes back to early work by \citet{Cher54a} (where it was called \emph{postulate~$5^*$}) and is also known as $\widehat\alpha$ \citep{BrHa11a}, the \emph{attention filter axiom} \citep{MNO12a}, and \emph{outcast}~\citep{AiAl95a}. The term \emph{strong superset property} was first used by \citet{Bord79a}. We refer to \citet{Monj08a} for a more thorough discussion of the origins of this condition.} 
Prominent Condorcet extensions that satisfy both SSP and monotonicity include
the \emph{Condorcet rule} (which selects a Condorcet winner whenever one exists and returns all alternatives otherwise), 
the \emph{top cycle} (also known as \emph{weak closure maximality}, \emph{GETCHA}, or the \emph{Smith set})
\citep{Good71a,Smit73a,Bord76a,Sen77a,Schw86a},
the \emph{minimal covering set}
\citep{Dutt90a}, 
the \emph{bipartisan set} \citep{LLL93b}
and variations of these \citep[see][]{Lasl97a,DuLa99a,Lasl00a,Bran11b}.\footnote{Remarkably, the robustness of the minimal covering set and the bipartisan set with respect to strategic manipulation also extends to \emph{agenda manipulation}. The strong superset property precisely states that an SCF is resistant to adding and deleting losing alternatives \citep[see also the discussion by][]{Bord83a}.
Moreover, both SCFs are composition-consistent, \ie they are strongly resistant to the introduction of clones \citep{LLL96a}. 
Scoring rules like plurality and Borda's rule are prone to both types of agenda manipulation \citep{Lasl96a,BrHa11a} as well as to strategic manipulation.}\footnote{Another prominent Condorcet extension---the \emph{tournament equilibrium set} \citep{Schw90a}---was conjectured to satisfy SSP and monotonicity for almost 20 years. This conjecture was recently disproved by \citet{BCK+11a}. In fact, it can be shown that the tournament equilibrium set as well as the related \emph{minimal extending set} \citep{Bran11b} can be Kelly-manipulated.} Two other SCFs that satisfy set-monotonicity are the \emph{Pareto rule} and the \emph{omninomination rule} (which returns all alternatives that are top-ranked by at least one agent).
\end{remark}

\begin{remark}[Coarsenings of Kelly-strategyproof SCFs]
For two SCFs $f,f'$, we say that $f$ is a \emph{coarsening} of $f'$ if $f'(R)\subseteq f(R)$ for all $R\in\mathcal{D}$. Kelly's preference extension has the useful property that $X \mathrel{\widehat{P}_i} Y$ implies $X' \mathrel{\widehat{P}_i} Y'$ for all non-singleton subsets $X'\subseteq X$ and $Y'\subseteq Y$. Hence, every coarsening $f$ of a Kelly-strategyproof SCF $f'$ is Kelly-strategyproof if $f(R)=f'(R)$ whenever $|f'(R)|=1$. As a consequence, the (McKelvey) \emph{uncovered set} \citep[see][]{Dugg11a}, a coarsening of the minimal covering set that returns singletons if and only if there is a Condorcet winner, is Kelly-strategyproof, even though it violates set-monotonicity.\footnote{For generalized strategyproofness as defined by \citet{Nehr00a} (see \remref{rem:weakprefs}), the second condition is not required and every coarsening of a strategyproof SCF is strategyproof.} In light of these comments, it seems interesting to try to identify \emph{inclusion-minimal} Kelly-strategyproof SCFs.
\end{remark}

\begin{remark}[Group-strategyproofness]
The proof of \thmref{thm:main} can be straightforwardly extended to show that no \emph{group} of agents can misstate their preferences in order to obtain a more preferred outcome.
\end{remark}

\begin{remark}[Fishburn-strategyproofness]
 It has been shown in other work that \thmref{thm:main} does not carry over to slightly more complete set extensions due to Fishburn and G\"ardenfors \citep{BrBr11a}. In fact, Pareto-optimality and Fishburn-strategyproofness are already incompatible within the class of majoritarian SCFs \citep{BrGe14a}.
\end{remark}

\begin{remark}[Necessary conditions]
It seems like there are no natural necessary conditions for Kelly-strategyproofness as long as preferences are linear. For example, any SCF $f$ such that $|f(R)|>(|A|/2)+1$ for all $R\in\mathcal{D}$ satisfies Kelly-strategyproofness simply because no pair of resulting choice sets is comparable. This observation allows one to easily construct Kelly-strategyproof SCFs that violate set-monotonicity, Maskin-monotonicity, or any other reasonable form of monotonicity.
\end{remark}

\begin{remark}[Weak preferences]\label{rem:weakprefs}
When individual preference relations do not have to be antisymmetric, a number of results mentioned in the introduction rule out the possibility of reasonable Kelly-strategyproof SCFs. A new result of this kind, which strengthens some existing theorems, is given as \thmref{thm:nocond} in the Appendix.

The proof of \thmref{thm:main} can be adapted for weak preferences to show that set-monotonicity implies Nehring's \emph{generalized strategyproofness}, a weakening of Kelly-strategyproofness \citep{Nehr00a}.\footnote{For generalized strategyproofness, it would also suffice to require a weakening of set-monotonicity in which the choice set can only get smaller when unchosen alternatives are weakened.} 
Generalized strategyproofness is defined by letting agent $i$ prefer set $X$ to set $Y$ if and only if $x \mathrel{P_i} y$ for all $x\in X$ and $y\in Y$.
\end{remark}

\begin{remark}[Matching markets]
\thmref{thm:main} can be straightforwardly extended to domains in which the indifference relation is fixed for each agent. This is, for example, the case in matching markets where each agent is typically assumed to be indifferent between all matchings in which his assignment is identical.
\end{remark}

\begin{remark}[Connected domains]
\thmref{thm:main} does not hold for a weakening of comprehensiveness, which \citet{Nehr00a} refers to as \emph{connectedness}. A domain $\mathcal{D}$ is connected if for all $i\in N$, $R_i,R_i'\in \mathcal{D}_i$, the following holds: if there is some $R_i'' \in \mathcal{L}\setminus\{R_i,R_i'\}$ with $R\cap R'\subseteq R''$, then there is some $R_i''\in\mathcal{D}_i\setminus\{R_i,R_i'\}$ with $R\cap R'\subseteq R''$. The following single-agent SCF $f$, defined on a connected---but not comprehensive---domain of size three, has been adapted from \citet{Nehr00a} and satisfies set-monotonicity while it violates Kelly-strategyproofness:
\[
\begin{array}{rll}
R_1 \colon &a, b, c, d\text{,} & \quad f((R_1))=\{c\}\text{,}\\
R'_1 \colon &a, b, d, c\text{,} & \quad f((R'_1))=\{a,b,c,d\}\text{,}\\
R''_1 \colon &b, a, d, c\text{,} & \quad f((R''_1))=\{b\}\text{.}
\end{array}
\]
\vspace{-2ex}
\end{remark}

\begin{remark}[Strategic abstention]
\citet{BrFi83a} introduced a particularly natural variant of strategic manipulation where agents obtain a more preferred outcome by \emph{abstaining} the election. \citet{Moul88b} has shown that every resolute Condorcet extension can be manipulated in this way and thus suffers from the so-called \emph{no-show paradox}. However, \citeauthor{Moul88b}'s proof strongly relies on resoluteness. If preferences over sets are given by Kelly's extension, set-valued Condorcet extension that satisfy Kelly-strategyproofness also cannot be manipulated by abstention under very mild conditions. This is, for instance, the case for all SCFs whose outcome only depends on pairwise majority margins (which covers all of the Condorcet extensions mentioned in Remark~\ref{rem:smon}) because any manipulation by abstention can be turned into a manipulation by strategic misrepresentation.\footnote{\citet{Pere01a} and \citet{JPG09a} have proved versions of \citeauthor{Moul88b}'s theorem \citep{Moul88b} for set-valued Condorcet extensions by using stronger assumptions on preference over sets.}

\end{remark}


%
%
%

\section*{Acknowledgments}

I am grateful to Florian Brandl, Markus Brill, and Paul Harrenstein for helpful discussions and comments. 
	This material is based on work supported by the Deutsche Forschungsgemeinschaft under grants \mbox{BR~2312/3-3}, \mbox{BR~2312/7-1}, and \mbox{BR~2312/7-2}. Early results of this paper were presented at the 22nd International Joint Conference on Artificial Intelligence (Barcelona, July 2011). A previous version of this paper, titled ``Group-Strategyproof Irresolute Social Choice Functions,'' circulated since 2010.

\appendix

\section{Appendix}

For the domain of transitive and complete (but not necessarily antisymmetric) preference relations $\mathcal{R}^N$, we show that all Condorcet extensions are Kelly-manipulable. This strengthens Theorem~3 by \citet{Gard76a} and Theorem~8.1.2 by \citet{Tayl05a}, who showed the same statement for a weaker notion of manipulability and a weaker notion of Condorcet winners, respectively. When assuming that pairwise choices are made according to majority rule, this also strengthens Theorems~1 and~2 by \citet{MaPa81a}.
However, our construction requires that the number of agents is linear in the number of alternatives.

\begin{theorem}\label{thm:nocond}
No Condorcet extension is Kelly-strategyproof in domain $\mathcal{R}^N$ when there are more than two alternatives.
\end{theorem}
\begin{proof}
Let $A=\{a_1,\dots,a_m\}$ with $m\ge 3$ and consider the following preference profile~$R$ with~$3m$ agents. In the representation below, sets denote indifference classes of the agents. 

\[
\begin{array}{rl}
R_1,R_2 \colon & \{a_2,\dots, a_m\}, a_1\\
R_3, R_4 \colon & \{a_1,a_3,\dots, a_m\}, a_2\\
\vdots\quad\;\;\ & \qquad\qquad\vdots\\
R_{2m-1}, R_{2m} \colon & \{a_1,\dots, a_{m-1}\}, a_m\\[1em]
R_{2m+1} \colon & \{a_3,\dots, a_m\}, a_1, a_2\\
\vdots\quad\;\;\ & \qquad\qquad\vdots\\
R_{3m-1} \colon & \{a_1,\dots, a_{m-2}\}, a_{m-1}, a_m\\
R_{3m} \colon & \{a_2,\dots, a_{m-1}\}, a_m, a_1
\end{array}
\]
For every alternative $a_i$, there are two agents who prefer every alternative to~$a_i$ and are otherwise indifferent. Moreover, for every alternative $a_i$ there is one agent who prefers every alternative except $a_{i+1}$ to $a_i$, ranks $a_{i+1}$ below $a_i$, and is otherwise indifferent.

Since $f(R)$ yields a nonempty choice set, there has to be some $a_i\in f(R)$. Due to the symmetry of the preference profile, we may assume without loss of generality that $a_2\in f(R)$.
Now, let 
\[\begin{array}{rl}
R_3',R_4' \colon & a_1, \{a_3,\dots,a_m\}, a_2
\end{array}\]
and define $R'=(R_{-3},R'_3)$ and $R''=(R'_{-4},R'_4)$. 
That is, $R'$ is identical to $R$, except that agent 3 lifted $a_1$ on top and $R''$ is identical to~$R'$, except that agent 4 lifted $a_1$ on top. Observe that $f(R'')=\{a_1\}$ because $a_1$ is the Condorcet winner in $R''$.

In case that $a_2\not\in f(R')$, agent 3 can manipulate as follows. Suppose $R$ is the true preference profile. Then, the least favorable alternative of agent~3 is chosen (possibly among other alternatives). He can misstate his preferences as in $R'$ such that~$a_2$ is not chosen. Since he is indifferent between all other alternatives, $f(R') \mathrel{\widehat P_3} f(R)$.

If $a_2\in f(R')$, agent~4 can manipulate similarly. Suppose~$R'$ is the true preference profile. Again, the least favorable alternative of agent~4 is chosen. By misstating his preferences as in $R''$, he can assure that one of his preferred alternatives, namely~$a_1$, is selected exclusively because it is the Condorcet winner in $R''$. Hence, $f(R'') \mathrel{\widehat P'_4} f(R')$.
\end{proof}

\end{document}